\documentstyle[12pt]{article}

\begin{document}
\thispagestyle{empty}

\begin{center}
\LARGE \tt \bf{ Distributional torsion of cosmic walls crossed by cosmic strings.}
\end{center}

\vspace{1cm}

\begin{center} {\large L.C. Garcia de Andrade\footnote{Departamento de
F\'{\i}sica Te\'{o}rica - Instituto de F\'{\i}sica - UERJ

Rua S\~{a}o Fco. Xavier 524, Rio de Janeiro, RJ

Maracan\~{a}, CEP:20550-003 , Brasil.

E-Mail.: GARCIA@SYMBCOMP.UERJ.BR}}
\end{center}

\vspace{1.0cm}

\begin{abstract}
Distributional sources of cosmic walls crossed by cosmic strings from Riemann-Cartan (RC) Geometry. The matter density of the planar wall is maximum at the point where the cosmic string crosses the cosmic wall. Cartan torsion is has a support on the cosmic string given by the Dirac $ \delta $-function. Off the sources are left with a torsionless vacuum.

\end{abstract}      
\vspace{1.0cm}       
\begin{center}
\Large{PACS number(s) : 0420,0450,1127}
\end{center}

\newpage
\pagestyle{myheadings}
\markright{\underline{Distributional torsion of cosmic walls crossed by cosmic strings.}}

\paragraph*{}
Recently Bezerra and Letelier \cite{1,2} have considered holonomy transformations to detect the presence of space-time defects \cite{3} such as cosmic strings and domain walls. In one of their examples \cite{2} they considered a topological  plane crossed by a cosmic string in the context of General Relativity (GR) where the space-time metric was given in terms of coordinates $ (t,z,{\rho},{\varphi}) $ by \cite{9}

\begin{equation} 
ds^{2}=e^{F}(dt^{2} + dz^{2})- {\rho}^{-8 {\lambda}} e^{H} ( d{\rho}^{2} + {\rho}^{2} d{\varphi}^{2})
\label{1}
\end{equation}

Where F and H depends on variable z orthogonal to the wall and $ {\lambda} $ is the mass density the cosmic string.

According to Cartan's calculus of exterior differential forms, metric (\ref{1}) can be reexpressed

\begin{equation}
\begin{array}{llll}
{\omega}^{0}= e^{\frac{F}{2}} dt  \nonumber \\
\\
{\omega}^{1}= e^{\frac{F}{2}} dz  \nonumber \\
\\
{\omega}^{2}= e^{\frac{H}{2}} {\rho}^{-4{\lambda}} d{\rho}  \nonumber \\
\\
{\omega}^{3}= e^{\frac{H}{2}} {\rho}^{-(4{\lambda}-1)} d{\varphi}    \\
\end{array}
\label{2}
\end{equation}
as

\begin{equation}
ds^{2} = {\eta}_{ab} {\omega}^{a} {\omega}^{b}
\label{3}
\end{equation}
where $ {\eta}_{ab} = diag(1,-1,-1,-1) $ is the Minkowski tetrad metric. Also recently several types of \cite{4,5} torsion defects have been considered by Letelier and myself \cite{6,7,8} following investigations on Riemannian space-time defects \cite{9}. These two facts together motivate us to investigate in this letter the role of torsion on the problem of cosmic torsion walls crossed by cosmic torsion strings as a new example of a mixed torsion defect. This topological torsion defect is obtained as a solution of (EC) equations.

Let us choose the following form for the 2-forms torsion

\begin{equation}
T^{1} = J^{1} {\omega}^{1} \wedge {\omega}^{0}
\label{4}
\end{equation}
where $ {{T}^{1}}_{10} = J^{1}, {{T}^{2}}_{20} = J^{2}$ and $ {{T}^{3}}_{30} = J^{3} $  are in principle the only nonvanishing components of the Cartan torsion tensor. From choice and the first Cartan's structure equation

\begin{equation}
T^{a} = d{\omega}^{a} + {{\omega}^{a}}_{b} \wedge {\omega}^{b}
\label{5}
\end{equation}
one obtains the following connection 1-forms

\begin{equation}
\begin{array}{llllll}
{{\omega}^{1}}_{0} = J^{1} {\omega}^{1} \nonumber \\
\\
{{\omega}^{2}}_{0} = J^{2} {\omega}^{2} \nonumber \\
\\
{{\omega}^{3}}_{0} = J^{3} {\omega}^{3} \nonumber \\
\\
{{\omega}^{3}}_{1} = \frac{H_{z}}{2} e^{\frac{-F}{2}} {\omega}^{3} \nonumber \\
\\
{{\omega}^{2}}_{1} = \frac{H_{z}}{2} e^{\frac{-F}{2}} {\omega}^{2} \nonumber \\
\\
{{\omega}^{3}}_{2} = (1-4{\lambda}) {\rho}^{4{\lambda}-1} e^{\frac{-F}{2}} {\omega}^{3} \nonumber
\end{array}
\label{6}
\end{equation}

By substitution of expressions (\ref{6}) into the Cartan's second eqn.

\begin{equation} 
{{R}^{a}}_{b} \equiv {{R}^{a}}_{bcd} ({\Gamma}) {\omega}^{a} \wedge {\omega}^{d} = d{{\omega}^{a}}_{b} + {{\omega}^{a}}_{c} \wedge {{\omega}^{c}}_{b}
\label{7}
\end{equation}
Riemann-Cartan (RC) curvature tensor components $ {{R}^{a}}_{bcd}({\Gamma}) $ read

\begin{equation}
\begin{array}{lllll}
{{R}^{2}}_{112} = \frac{1}{2}[H_{zz} - \frac{H_{z}F_{z}}{2} + \frac{{H_{z}}^{2}}{2}]e^{-F} \nonumber \\
\\
{{R}^{3}}_{213} = - (1-4{\lambda}) {\rho}^{4{\lambda}-1} \frac{H_{z}}{2}e^{\frac{-F}{2}} \nonumber \\
\\
{{R}^{3}}_{232} = \frac{{H_{z}}^{2}}{4} e^{-F} + (1-4{\lambda}) H_{z}  {\rho}^{4{\lambda}-1} e^{-\frac{(H+F)}{2}} \nonumber \\
\\
{{R}^{1}}_{021} = J^{1}_{\rho} {\rho}^{4{\lambda}} e^{-\frac{H}{2}} \nonumber \\
\\
{{R}^{3}}_{031} = {{R}^{2}}_{021} = \frac{H_{z}}{2} J^{1} e^{-\frac{H}{2}} \nonumber
\end{array}
\label{8}
\end{equation}
where to simplify matters we have considered $ J^{1} $ as the only surviving component of torsion .

By making use of the definition of a topological defect which states that the curvature $ {{R}^{a}}_{bcd} ({\Gamma}) $ must vanish off the space-time defect one obtains

\begin{equation}
\begin{array}{ll}
J^{1}_{\rho} = 0 \nonumber \\
\\
H_{z} = 0 \nonumber
\end{array}
\label{9}
\end{equation}

Here $ H_{z} =0 $ tell us that $ H (z) = const. $. Thus off the defects torsion vanishes and the metric is flat if one considers $ F(z) = H(z) $. Notice also that we have assumed that $ J^{1} = J^{1}({\rho}) $ and not on Z. This means physically that the torsion is linked with the string singularity $ {\rho} = 0 $ . Therefore we are led to define $ J^{1}({\rho}) \equiv {\lambda} {\delta}({\rho}) $.

By contraction of expressions (\ref{8}) one obtains the Ricci and Einstein tensor $ R_{ab} $ and $ G_{ab} $.

\begin{equation}
\begin{array}{lll}
R_{11} = \frac{1}{2} [H_{zz} - \frac{H_{z} F_{z}}{2} + \frac{{{H}_{z}}^{2}}{2}] e^{-F} \nonumber \\
\\
R_{22} = - \frac{{H_{z}}^{2}}{4} e^{-F} + (1-4{\lambda}) H_{z} e^{-(\frac{H+F}{2})} {\rho}^{-(1-4{\lambda})} \nonumber \\
\\
R_{33} = \frac{{H_{z}}^{2}}{4} e^{-F} + (1-4{\lambda}) H_{z}{\rho}^{-(1-4{\lambda})} e^{-(\frac{H+F}{2})} \nonumber
\end{array}
\label{10}
\end{equation}
where $ R=-(R_{11}+R_{22}+R_{33}) $ is the Ricci scalar.

The Einstein tensor is
\begin{equation}
\begin{array}{lll}
G_{00} = - \frac{1}{2} [ H_{zz} + (1-4{\lambda}) H_{z} {\rho}^{4{\lambda}} ] = T^{(w)}_{00} + T^{(cs)}_{00} \nonumber \\
\\
G_{11} = \frac{1}{2} (1-4{\lambda}) H_{z} {\rho}^{-(1-4{\lambda})} = T^{(cs)}_{11} \nonumber \\
\\
G_{22} = G_{33} = \frac{1}{2} H_{zz} = T^{(w)}_{22} = T^{(w)}_{33} \nonumber
\end{array}
\label{11}
\end{equation}
where in (\ref{11}) we have already considered the linear approximation dropping terms like $ H_{z}F_{z} ,H^{2}_{z} $ etc.  Besides to simplify matters we also consider that $ F(z) \equiv H(z) $. Therefore the solution reads

\begin{equation}
\begin{array}{lll}
\frac{1}{2} H_{zz} = \frac{{\sigma}}{z} {\delta}(z) \equiv {\rho}_{w} \nonumber \\
\\
\frac{1}{2}{(1-4{\lambda})}H_{z} {\rho}^{-(1-4{\lambda})} = {\rho}_{cs} \nonumber \\
\\
R_{21} = -(1-4{\lambda}) {\rho}^{4{\lambda}-1} \frac{H_{z}}{2} e^{-\frac{H}{2}} \nonumber
\end{array}
\label{12}
\end{equation}
where the stress-energy tensor of the system cosmic wall plus cosmic torsion string is given by 

\begin{equation}
\begin{array}{ll}
\stackrel{(w)}{T}^{a}_{b} = {\rho}_{w} \mbox{diag(1,0,1,1)} \nonumber \\
\\
\stackrel{(cs)}{T}^{a}_{b} = {\rho}_{cs} \mbox{diag(1,1,0,0)} \nonumber
\end{array}
\label{13}
\end{equation}

From (\ref{12}) and (\ref{13}) we notice that the energy-momentum distribution describes a wall where the density is not constant to the presence of cosmic string at $ z = 0 $. Notice that the effective wall density is given by $ {\sigma}^{eff} = \frac{\sigma}{z} $.

The remaining field equations contain the torsion contribution and they are

\begin{equation}
\begin{array}{lll}
R_{02}({\Gamma}) = J^{1}_{\rho} {\rho}^{4{\lambda}} e^{-\frac{H}{2}} \nonumber \\
\\
R_{01}({\Gamma}) = - H_{z} J^{1} e^{-\frac{H}{2}} \nonumber
\end{array}
\label{14}
\end{equation}
since $ J^{1} = {\lambda} {\delta}({\rho}) $ thus 

\begin{equation}
\begin{array}{ll}
R_{02}({\Gamma}) = - {\lambda} {\delta}({\rho}) {\rho}^{4{\lambda}-1} e^{-\frac{H}{2}} \nonumber \\
\\
R_{01}({\Gamma}) = - {\lambda} {\delta}(z){\delta}({\rho})e^{-\frac{H}{2}} \nonumber
\end{array}
\label{15}
\end{equation}
are the skew-symmetric parts of the Ricci tensor in RC-space. From expression one may infer that

\begin{equation}
H(z) = {\sigma} {\theta}(z) + b
\label{16}
\end{equation}
where b are integration constants. Thus the metric (\ref{1}) reads

\begin{equation}
ds^{2} = e^{{\sigma}{\theta}(z)+b} [(dt^{2} - dz^{2}) - {\rho}^{-8 {\lambda}}(d{\rho}^{2}+{\rho}^{2} d{\varphi}^{2})]
\label{17}
\end{equation}
Therefore in our example torsion only appears in the off-diagonal components of the Ricci tensor.

Note from (\ref{17}) that in the absence of cosmic string $({\lambda}=0)$ the space is flat on both sides of the wall. Another interesting example is to consider the interaction between the spinning cosmic string and cosmic walls. Finally are would like to mention that what Baker \cite{9} calls torsion string is a concept which differs of ours in the sense that his torsion is not a $ {\delta}$-Dirac.

\section*{Acknowledgments}
\paragraph*{}
Thanks are due to Prof. P.S.Letelier. for his constant advice on the subject of this paper. I am veryt much indebt to Universidade do Estado do Rio de Janeiro(UERJ) and CNPq. (Brazilian Government Agency) for financial support.


\begin{thebibliography}{9}
\bibitem{1} V.Bezerra and P.S.Letelier, (1991), Class. Quant. Grav., \underline{8},L73.
\bibitem{2} V.Bezerra and P.S.Letelier, (1997), J. Math. Phys.,(37)12,6271.
\bibitem{3} P.S.Letelier, (1995), Class. Quant. Grav., \underline{14}, L75.
\bibitem{4} P.S.Letelier, (1995), Class. Quant. Grav., 2221.
\bibitem{5} P.S.Letelier, (1995), Class. Quant. Grav., 12,471.
\bibitem{6}L.C.Garcia de Andrade, (1997), On non-Riemannian domain walls., DFT-UERJInternal Reports.
\bibitem{7}L.C.Garcia de Andrade, (1998), J.Math. Phys., (39).
\bibitem{8}L.C.Garcia de Andrade, (1998), Modern. Phys. Lett.A ,\underline{12} ,27 ,2005.
\bibitem{9} W.M.Baker, Class. Quant.Grav., (1990),\underline{7},717.
\end{thebibliography}
\end{document}